\documentclass[final,authoryear,5p,times,twocolumn]{elsarticle}
\usepackage{graphicx}
\usepackage{amssymb}
\usepackage{amsmath}

\journal{New Astronomy Reviews}

\begin{document}

\newcommand\apj{\rmfamily{ApJ}}%
\newcommand\apjl{\rmfamily{ApJ}}%
\newcommand\apjs{\rmfamily{ApJS}}%
\newcommand\aap{\rmfamily{A\&A}}%
\newcommand\aapr{\rmfamily{A\&A~Rev.}}%
\newcommand\aaps{\rmfamily{A\&AS}}%
\newcommand\araa{\rmfamily{ARA\&A}}%
\newcommand\memsai{\rmfamily{Mem.~Soc.~Astron.~Italiana}}%
\newcommand\mnras{\rmfamily{MNRAS}}%
\newcommand\na{\rmfamily{New A}}%
\newcommand\nat{\rmfamily{Nature}}%
\newcommand\pasj{\rmfamily{PASJ}}%
\newcommand\pasp{\rmfamily{PASP}}%
\newcommand\physrep{\rmfamily{Phys.~Rep.}}%
\newcommand\ssr{\rmfamily{Space~Sci.~Rev.}}%
\let\astap=\aap
\let\apjlett=\apjl
\let\apjsupp=\apjs

\begin{frontmatter}

\title{The broad Fe K$\alpha$ line and supermassive black holes}

\author[aob]{Predrag Jovanovi\'{c}\corref{cor}}
\ead{pjovanovic@aob.bg.ac.rs}
\address[aob]{Astronomical Observatory, Volgina 7, 11060
Belgrade, Serbia}
\cortext[cor]{Corresponding author}

\begin{abstract}
Here we present an overview of some of the most significant
observational and theoretical studies of the broad Fe K$\alpha$
spectral line, which is believed to originate from the innermost
regions of relativistic accretion disks around central supermassive
black holes of galaxies. The most important results of our
investigations in this field are also listed. All these
investigations indicate that
the broad Fe K$\alpha$ line is a powerful tool for studying the
properties of the supermassive black holes (such as their masses and
spins), space-time geometry (metric) in their vicinity, their
accretion physics, probing the effects of their strong
gravitational fields, and for testing the certain predictions of
General Relativity.
\end{abstract}


\end{frontmatter}

\section{Supermassive black holes}

According to General Relativity \citep{ein16}, a black hole
\citep[named by][]{whe68} is a region of space-time around a
collapsed mass with a gravitational field that has became so strong
that nothing (including electromagnetic radiation) can escape from
its attraction, after crossing its event horizon \citep{wald84}.
The problem of such catastrophic gravitational collapse was first
adressed by \citet{chandra31} when he discovered the upper mass limit
for ideal white dwarfs, composed of a degenerate electron-gas.
Subsequent studies by \citet{lan32,opp39} and \citet{pen65}
established the modern theory of gravitational collapse. The theory
of black holes, as well as their fundamental
properties are presented in numerous monographs and papers
\citep[see e.g.][]{car73,chandra83}, and therefore in this review we
will focus only on those studies which are necessary for the
discussion of the relation between black holes and the Fe
K$\alpha$ spectral line.

All black holes in nature are commonly classified according to their
masses as: supermassive black holes (with masses $M_{BH}\sim
10^5-10^{10}\ M_\odot$), intermediate-mass black holes ($M_{BH}\sim
10^2-10^5\ M_\odot$), stellar-mass black holes ($M_{BH} < 10^2\
M_\odot$), mini and micro black holes ($M_{BH} \ll M_\odot$). A
crucial event for the acceptance of black holes was the discovery of
pulsars by \citet{hb68}, because it was the clear evidence of the
existence of neutron stars, and therefore, confirmation of
Chandrasekhar limit. The first detection of a solar mass black hole
came in 1972, when the mass of the rapidly variable X-ray source
Cygnus X-1 was proven to be above the maximum allowed for a neutron
star \citep[see e.g.][and references therein]{ff05}.

Nowadays, it is widely accepted that supermassive black holes are
located in the centers of most of galaxies, and thus have a
fundamental influence on galactic formation and evolution. Some of
the first indirect arguments that they exist in galactic nuclei are
given by \citet{lyn69} and \citet{lyn71}. Also, according to the
unification model of active galactic nuclei -- AGN
\citep{ant93,pet97,krol99}, they are most likely powered by the
accretion of gas onto their central supermassive black holes with
mass ranging from $10^5$ to $10^9\ M_\odot$ \citep{Kaspi00,pet04}.

\subsection{Space-time geometry in vicinity of supermassive black
holes}

In general, black holes have three measurable parameters (not
including the Hawking temperature): charge, mass (and hence
gravitational field) and angular momentum (or spin). In the case of
supermassive black holes, only the latter two are of sufficient
importance because they are responsible for several effects which
can be observationally detected \citep[for more details see
e.g.][]{jov08}.

A central supermassive black hole of a galaxy can be stationary or
rotating. In the first case, space-time geometry in the black hole
vicinity depends only on its mass $M$ and is described by the
Schwarzschild metric \citep{schw16}, while in the latter case,
space-time geometry depends also on spin $a$ (i.e. angular momentum
$J=a M$) of the black hole and is described by Kerr metric
\citep{kerr63}. Therefore, Schwarzschild metric describes the
spherically symmetric gravitational field around a time-steady
non-rotating black hole in vacuum, while the Kerr metric describes
the gravitational field outside an uncharged rotating black hole
and, contrary to the Schwarzschild metric, is no longer spherically
symmetric. \citet{bl67} discovered a generalized coordinate system
in which Kerr metric is most commonly used today \citep[see
e.g.][]{car73}.

Several characteristic radii can be defined around black holes, and
the most important are \citep[see e.g.][]{jov09}: 1. Schwarzschild
radius $R_S=2GM/c^2$ (representing the limiting radius below which a
collapsed mass form a spherically symmetric non-rotating black
hole), where $G$ is the gravitational constant and $c$ is the speed
of light; 2. gravitational radius $R_g$ (being a half of $R_S$ and
usually used as a unit for distance around a black hole); 3. radius
of event horizon $R_h$ (representing space-time boundary below which
events cannot affect an outside observer); and 4. radius of
marginally stable orbit $R_{ms}$ (representing the minimum allowed
radius of a stable circular equatorial orbit around a black hole).

\begin{figure}[ht!]
\centering
\includegraphics[width=\columnwidth]{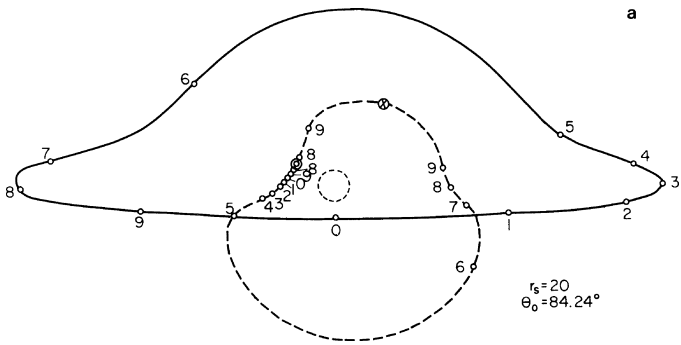}
\caption{A circular orbit (solid line) of a point source in the
equatorial plane around an extreme Kerr metric black hole.
Image Credit: Cunningham \& Bardeen, Copyright: ApJ, 183, 237 (1973).}
\label{fig01}
\end{figure}

The equations governing photon trajectories in the Kerr metric are
derived by \citet{car68}. \citet{mtw73} showed that certain
relativistic effects in vicinity of a black hole, such as light
banding, gravitational and Doppler shifts as seen by distant
observer could be computed by solving the equation of the photon
orbit. \citet{cun72,cun73} calculated radiation of a point source in
a circular orbit in the equatorial plane around an extreme Kerr
metric black hole (see Fig. \ref{fig01}).

\citet{bard70} studied the effects of accretion from a disk of gas
orbiting a black hole on its properties, and showed that the Kerr
metric is more appropriate for describing supermassive black holes
than the Schwarzschild metric. This conclusion is strengthened even
more by
\citet{th74}, who showed that a black hole at the center of an
accretion disk would be spun up to the maximum possible value of $a
\approx 0.998$ in a relatively short time. In such a case the disk
would be extended down to about 1.23 $R_g$, while in the case of a
non-rotating black hole it could extend only down to 6 $R_g$.

\begin{figure*}[ht!]
\centering
\includegraphics[width=\textwidth]{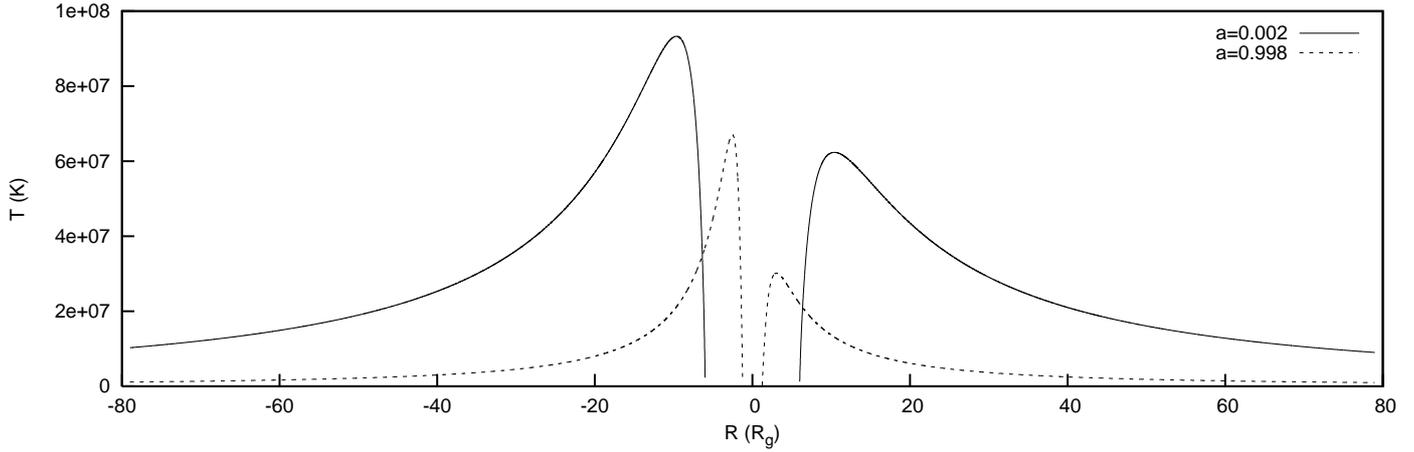}
\caption{The distribution of the temperature as a function of the
disk radius $R$, given for two different values of angular momentum
$a$. Negative values of $R$ correspond to the approaching and
positive values to the receding side of the disk. Image Credit:
Popovi\'{c} et al., Copyright: ApJ, 637, 620 (2006).} \label{fig02}
\end{figure*}

\begin{figure*}[ht!]
\centering
\includegraphics[width=0.9\textwidth]{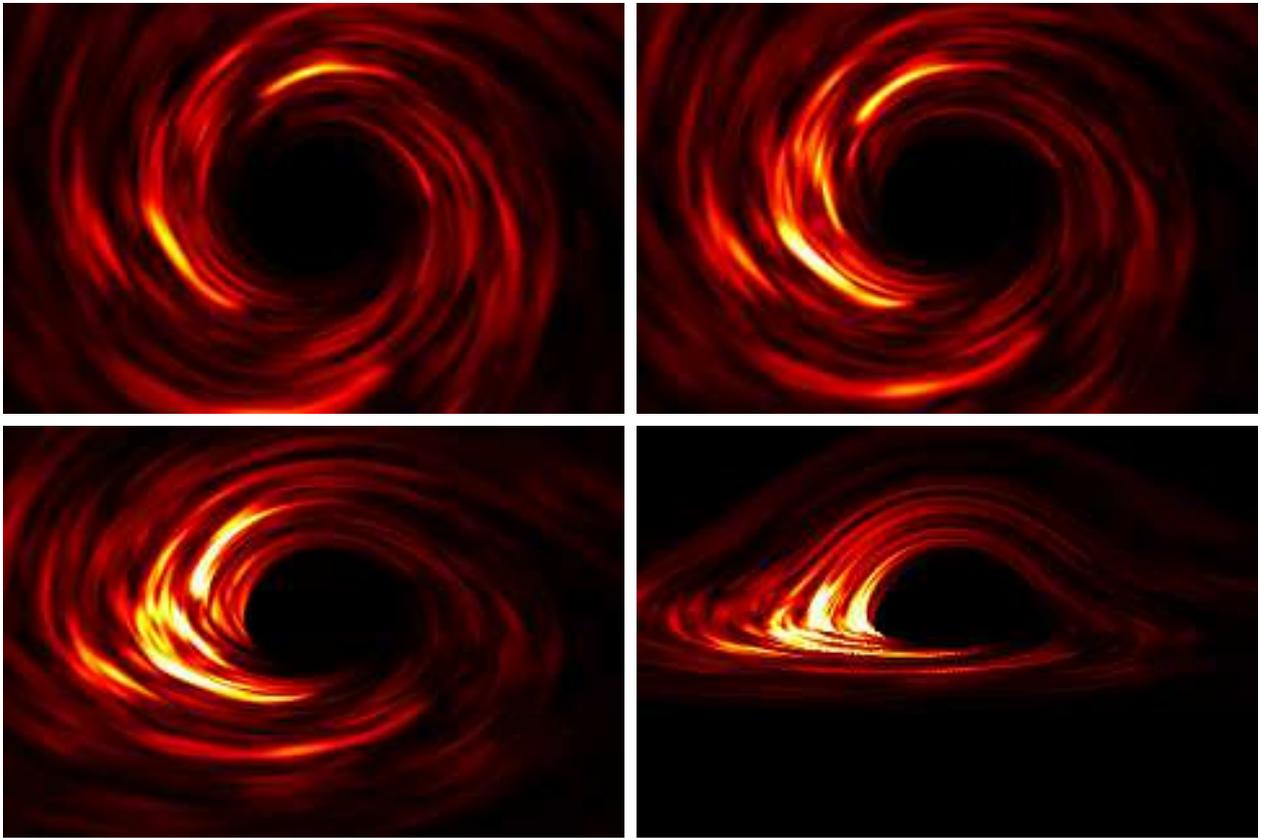}
\caption{Magnetohydrodynamic simulations of an accretion disk as
seen by a distant observer at the following inclination angles
\citep{Armitage03}: 5$^\circ$ (upper left), 30$^\circ$ (upper
right), 55$^\circ$ (lower left) and 80$^\circ$ (lower right). The
disk rotation is in counterclockwise direction. Image Credit:
Armitage \& Reynolds, Copyright: MNRAS, 341, 1041 (2003).}
\label{fig03}
\end{figure*}

\subsection{Accretion onto the supermassive black holes}

Some observed quasars have luminosities of up to $L_{bol} \sim
10^{47}$ erg/s \citep{Schneider06}. The corresponding total energy
emitted during the lifetime of such quasars can be estimated to $E
\geq 3\times 10^{61}$ erg, assuming that their minimum age is about
$10^7$ years and that their luminosities  do not change
significantly over the lifetime \citep{Schneider06}. Accretion is the
only mechanism which can yield sufficient
efficiency $\epsilon$ (defined as the mass fraction of fuel that is
converted into energy, according to $E=\epsilon m c^2$) to
explain such high energies. The maximum efficiency of accretion
is about $\epsilon \sim 6\%$ for a non-rotating black hole and
$\epsilon \sim 29\%$ for a maximally rotating one
\citep{Schneider06}.

On the other hand, if the cold matter which was initially at rest
and without magnetic
field, was subjected to free radial infall it would accrete to the
central black hole without any energy release or observational
effects \citep{Shakura73}. However, in the case of AGN the accreting
matter has a significant angular momentum which does not allow its
free infall. At the marginally stable orbit of central black hole
with mass of $10^8M_\odot$, the specific angular momentum is
$\approx 1\times 10^{24}\ \mathrm{cm}^2/\mathrm{s}$, which is much
less even in comparison to a typical galaxy where the speciffic
angular momentum of orbiting material is $\approx 6\times 10^{28}\
\mathrm{cm}^2/\mathrm{s}$ \citep{krol99}. It means that approach
of accreting material toward a black hole requires a loss of the
greatest fraction of its initial angular momentum. Mechanisms which
can contribute to such loss of angular momentum are viscosity,
nonaxisymmetric gravitational forces, magnetic forces
\citep{krol99}, outflowing winds \citep{beg83,elv00}, or the
Magnetorotational Instability (MRI) which was discovered by
\citet{vel59} and \citet{chan60}, but did not come to the attention
of the astrophysical community until rediscovered by
\citet{balb91,balb98}.

Since the orbit of minimum energy for fixed angular momentum in any
spherically symmetric potential is a circle, the infall of accreting
material due to loss of its angular momentum will be in form of
successively smaller and smaller concentric circles. Matter
traveling along orbits inclined to each other will eventually
collide in the plane of intersection, and as a result, the angular
momenta of different gas steams will be mixed and equalized.
Consequently, all accreting matter will orbit in a single plane and
will have the same specific angular momentum at any given radius,
meaning that accretion is most likely performed through an accretion
disk \citep{krol99}. The assumption of a disk geometry for the
distribution of the X-ray emitters in the central parts of AGN is
also supported by the spectral shape of the observed Fe K$\alpha$
line \citep{Nandra97,Nandra99,Nandra07}.

\citet{Shakura73} developed what is now called ''standard model of
accretion disk'', in which accretion occurs via an optically thick
and geometrically thin disk and where the spectrum of thermal
radiation emitted from the disk surface depends on its structure and
temperature, and therefore on the distance to the central black
hole. The distribution of the temperature along the radius of the
disk in this model is presented in Fig. \ref{fig02} \citep[see][for
more details]{pop06}. This model was originally developed to
describe the accretion disks around stellar sized black holes in the
binary systems, but with certain modifications it could be also
applied on accretion disks around central supermassive black holes
of AGN.

The total energy release of emitted radiation is mainly determined
by the rate of matter inflow into the disk on its outer boundary,
i.e. by its accretion rate $\dot{M}$. If the accretion converts
matter to radiation with fixed radiative efficiency $\eta$ (in
rest-mass units), then a characteristic scale for the accretion rate
is the Eddington accretion rate for which the total release of
energy in the disk $L=\eta\dot{M}c^2$ is equal to the Eddington
luminosity, a critical luminosity beyond which the radiation force
overpowers gravity for any given mass \citep{krol99,Shakura73}.
Observed AGN have luminosities from $\sim 10^{42}$ to $\sim 10^{48}$
erg/s, which means that their central black holes must have masses
from $10^5$ to $10^9\ M_\odot$ \citep{krol99}.

Using the Faint Object
Spectrograph (FOS) on the Hubble Space Telescope (HST),
\citet{harm94} found strong evidence for a disk of ionized gas around
a supermassive nuclear black hole in M87.

\citet{Armitage03} performed magnetohydrodynamic simulations of an
accretion disk as seen by a distant observer at four different
inclinations (see Fig. \ref{fig03}). These simulations showed the
existence of especially bright regions in form of arcs within the
turbulent flow, which trace out the photon trajectories close to the
radius of marginally stable orbit. These arcs are much brighter on
approaching side of the disk due to higher surface temperature, as
shown in Fig. \ref{fig03}, and also due to Doppler boosting and
relativistic beaming which both enhance the flux observed in the
direction of motion and diminish the flux in the opposite direction.
The influence of these relativistic effects on the shape of the Fe
K$\alpha$ line emitted from the accretion disk will be explained in
more details in \S2.3.

\section{Fe K$\alpha$ spectral line}

\subsection{Observational studies}

Some observational studies suggest that the broad
fluorescent/recombination iron K$\alpha$ line at 6.4 keV originates
from the innermost part of accretion disk, close to the central
black hole. For example, using XMM-Newton observation of the
broad-line radio galaxy 4C 74.26, \citet{ball05} found a broad Fe
K$\alpha$ line emitted from a region which had inner radius close to
the innermost stable circular orbit ($R_{ms}$) for a maximally
spinning black hole and outer radius within 10 $R_g$. Therefore, the
broad Fe K$\alpha$ line is an important indicator of accreting flows
around supermassive black holes. At the same time, it is the
strongest line of the X-ray radiation, and it can be found in the
spectra of all types of accreting sources: binary black hole and
neutron star systems, cataclysmic variable stars and AGN.

\begin{figure}[ht!]
\centering
\includegraphics[width=0.9\columnwidth]{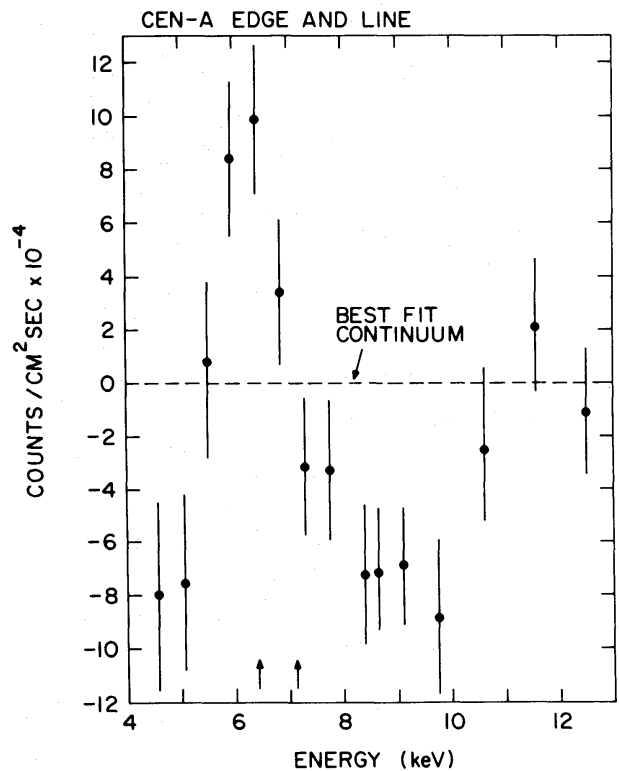}
\caption{The Fe K$\alpha$ line of Cen A (NGC 5128)
observed by OSO-8 during 1975 and 1976 \citep{mus78}.
Image Credit: Mushotzky et al., Copyright: ApJ, 220, 790 (1978).}
\label{fig04}
\end{figure}

It was first discovered by \citet{mus78} in OSO-8 X-ray observations
of Cen A (NGC 5128) during 1975 and 1976 (see Fig. \ref{fig04}). NGC
5128 is one of the closest radio galaxies and one of the first
extragalactic objects to be identified as an X-ray source. The
reflection of the X-rays in K lines of heavy elements from a cold
surface was also discussed in the case of the X-ray binaries
\citep[see e.g.][]{bas78}, and Exosat Observatory detected a broad
emission iron K line at 6.2 keV in the X-ray spectrum of Cyg X-1, a
binary X-ray source which was the best known black hole candidate
\citep{bar85}. Ginga satellite detected a strong iron K$\alpha$
fluorescence line in a number of Seyfert 1 galaxies \citep[see
e.g.][]{kun90,mm90,pir90,pou90,nan91}. The Fe K$\alpha$ line of
active galactic nuclei, as well as different geometries and
astrophysical conditions necessary for its emergence, was
intensively studied around 1990s \citep[see e.g.][and references
therein]{mat91,geo91}.

\begin{figure}[ht!]
\centering
\includegraphics[width=\columnwidth]{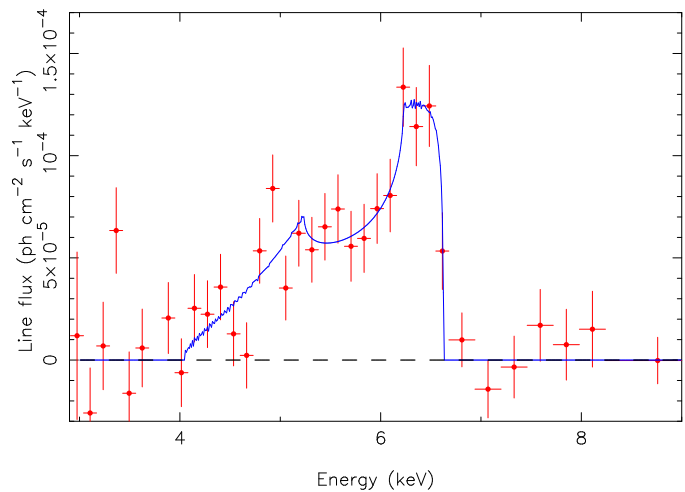}
\caption{The profile of the Fe K$\alpha$ line from Seyfert 1 galaxy
MCG-6-30-15 observed by
\emph{ASCA SIS} detector \citep{Tanaka95} and the best fit (blue
solid line) obtained by
a model of the accretion disk in Schwarzschild metric, extending
between 3 and 10 Schwarzschild radii \citep{fab89}.
Image Credit: Tanaka et al., Copyright: Nature, 375, 659 (1995).
Image generated by Dr. Paul Nandra NASA/GSFC.}
\label{fig05}
\end{figure}

The CCD detectors on Japanese \emph{ASCA} satellite were the first
instruments with sufficient spectral resolution and sensitivity in
the X-ray band, by which \citet{Tanaka95} obtained the first
convincing proof for the existence of a relativistically broadened Fe
K$\alpha$ line in AGN spectra. This discovery was made after four-day
observations of
Seyfert 1 galaxy MCG-6-30-15 (see Fig. \ref{fig05}).

Several studies have been performed over samples of local AGN
\citep[see
e.g.][etc]{Nandra97,Yaqoob05,Nandra07,Bianchi08,Markowitz08}, as
well as from distant quasars \citep{Corral08} in order to
characterize the Fe K$\alpha$ emission. \citet{Nandra07} performed a
spectral analysis of a sample of 26 type 1 to 1.9 Seyferts galaxies
($z < 0.05$) observed by \emph{XMM-Newton}. They found that a
relativistic line is significantly detected in a half of their
sample (54$\pm$10\%) with a mean equivalent width (EW) of
$\sim$~80~eV, but around 30\% of selected AGN showed a relativistic
broad line that can be explained by the emission of an accretion
disk.

Recent Chandra observations revealed a narrow component of the Fe
K$\alpha$ line in the X-ray spectra of many
AGN \citep[see e.g.][]{yaq01,yaq04,page04,shu10}. Although the origin
of this narrow component is still poorly understood, it is thought to
be produced by emission from material much further from the central
black hole, probably in the outermost regions of the accretion disk,
in the Broad Line Region (BLR) or torus. Therefore, in the further
text the focus will be only on the broad component of the Fe
K$\alpha$ line, which originates from the innermost parts of an
relativistic accretion disk.

If the Fe K$\alpha$ line originated from an arbitrary radius of a
non-relativistic (Keplerian) accretion disk it would have a
symmetrical profile (due to Doppler effect) with two peaks: a "blue"
one which is produced by emitting material from the approaching side
of the disk in respect to an observer, and a "red" one which
corresponds to emitting material from the receding side of the disk.
Broadening of the Fe K$\alpha$ line arises mostly from General
Relativistic effects and large rotational velocities of material
emitting near the black hole. Using \emph{ASCA} satellite
observations,
\citet{Nandra97} found that, in case of 14 Seyfert 1 galaxies,
Full-Widths at Half-Maximum (FWHM) of their Fe K$\alpha$ lines
correspond to velocities of $\approx 50,000$ km/s. In some cases
(e.g. for Seyfert 1 galaxy MCG-6-30-15), FWHM velocity reaches 30\%
of speed of light \citep[see e.g.][]{Nandra07}. It means that in the
vicinity of the central black hole, orbital velocities of the
emitting material are relativistic, causing the enhancement of the
Fe K$\alpha$ line ''blue'' peak in regard to its ''red'' peak
(relativistic beaming). Taking into account the integral emission in
the line over all radii of accretion disk, one can obtain the line
with asymmetrical and highly broadened profile \citep{fab00}. The
''blue'' peak is then very narrow and bright, while the ''red'' one
is wider and much fainter (see Fig. \ref{fig05}). Besides, the
gravitational redshift causes further deformations of the Fe
K$\alpha$ line profile by smearing the ''blue'' emission into
''red'' one (see \S2.3 for more details). Since the observed Fe
K$\alpha$ line profiles are strongly affected by such relativistic
effects,
they represent a fundamental tool for investigating the plasma
conditions and the space-time geometry in the vicinity of the
supermassive black holes of AGN.

\subsection{Production of the Fe K$\alpha$ line}

\begin{figure}[ht!]
\centering
\includegraphics[width=0.95\columnwidth]{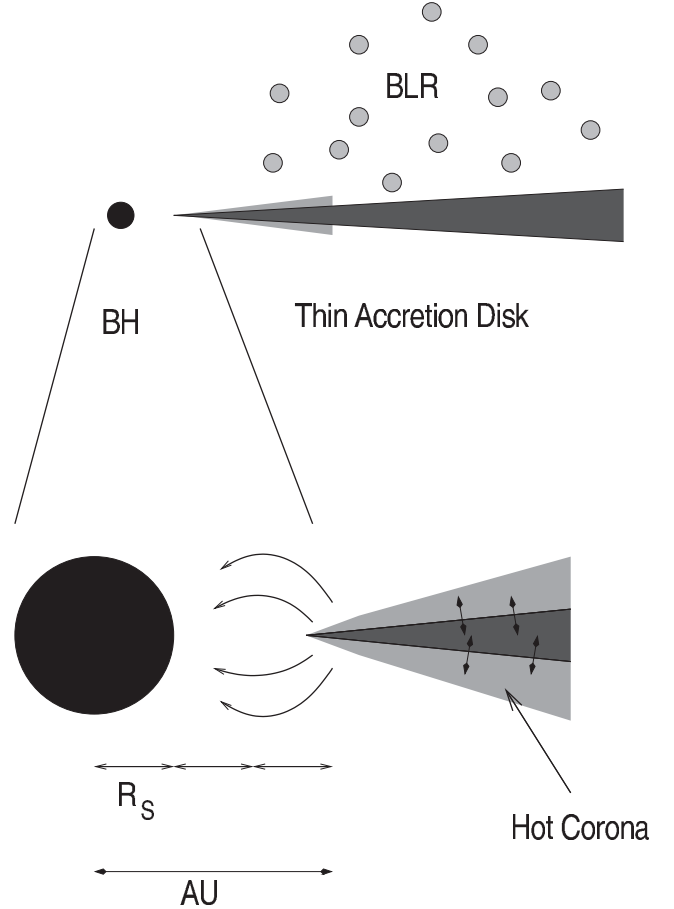}
\caption{Schematic illustration of inner region of accretion disk
around black hole, where the Fe K$\alpha$ line is produced.
Image Credit: Wilms et al., Copyright: MemSAI, 75, 519 (2004).}
\label{fig06}
\end{figure}

A substantial amount of X-ray radiation of AGN is thought to be
emitted from the hot corona sandwiching the accretion disk (see Fig.
\ref{fig06}). The Fe K$\alpha$ line is produced when the hot corona
irradiates the relatively cold accretion disk by the hard X-ray
power law continuum, causing among the rest, photoelectric
absorption followed by fluorescent line emission \citep{fab00}. When
plasma is subjected to the influence of the hard X-ray radiation,
one of the two $K$-shell ($n=1$, where $n$ is the principal quantum
number) electrons of an iron atom (or ion) is ejected following the
photoelectric absorption of an X-ray \citep{fab00}. The threshold
for the absorption by neutral iron is $7.1$ keV \citep{fab00}. The
resulting excited state decays when an $L$-shell ($n=2$) electron
drops into the $K$-shell, releasing $6.4$ keV of energy. This energy
is either emitted as an emission-line photon (34\% probability) or
internally absorbed by another electron (66\% probability) which is
consequently ejected from the iron ion (Auger effect).

The fluorescent yield (i.e. the probability that photoelectric
absorption is followed by fluorescent line emission rather than the
Auger effect) is a weak function of the ionization state from
neutral iron (Fe I) up to Fe XXIII \citep{fab00}. For
lithium-like iron (Fe XXIV) through to hydrogen-like iron (Fe XXVI),
the lack of at least two electrons in the $L$-shell means that the
Auger effect cannot occur. For He and H-like iron ions, the line is
produced by the capture of free electrons (recombination) and the
equivalent fluorescent yield is high and it depends on the plasma
conditions \citep{fab00}.

For the neutral iron, the Fe K$\alpha$ line energy is 6.4 keV (more
precisely, there are two components of the line \citep{fab00}: Fe
K$\alpha_1$ at 6.404 and Fe K$\alpha_2$ at 6.391 keV), while in the
case of ionization, the energy of both the photoelectric threshold
and the Fe K$\alpha$ line are slightly increased. Even for such high
ionization states of He and H-like iron ions, the Fe K$\alpha$ line
energy is increased only to $6.7$ and $6.9$ keV, respectively
\citep{krol99}. Importance of ionization, X-ray scattering and
Compton reflection on the properties of the observed Fe K$\alpha$
line, was discussed by \citet{tur97}, based upon the analysis of
\emph{ASCA} observations of 25 Seyfert 2 galaxies.

Fe K$\alpha$ line is pretty narrow in itself, but in case when it
originates from a relativistically rotating accretion disk of AGN it
becomes wider due to kinematic effects, and also its shape (or
profile) is changed due to Doppler boosting and gravitational
redshift (see Fig. \ref{fig07}). Such broadening of the line is very
often observed in spectra of Seyfert galaxies and is one of the main
evidences for the existence of a relativistic accretion disk which
extends deeply in the gravitational field of the central black hole
\citep{Zycki04}.

\subsection{Theoretical studies}

In general, there are several approaches to numerically evaluate the
line profiles emitted by the accretion disk around black hole
\citep{rey03}: i) analytically in the weak field limit
\citep{chen89b,chen89a} and in the Schwarzschild metric
\citep{fab89,ste90,mat92}, ii) by brute-force using the direct
integration of the photon trajectory in the Kerr metric
\citep[e.g.][]{ka92}, iii) by expressing observed flux over
transfer-function containing all relativistic effects \citep{cun75},
and iv) using so called ray-tracing method in Kerr metric
\citep{Bao94,Bromley97,Fanton97,Cadez98}.

\begin{figure}[ht!]
\centering
\includegraphics[width=0.98\columnwidth]{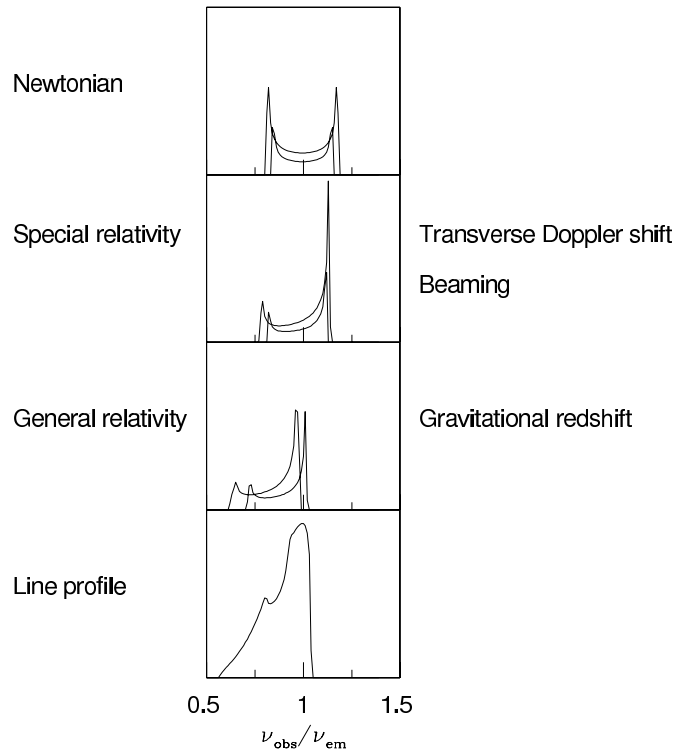}
\caption{The influence of Doppler and transverse-Doppler shifts,
relativistic beaming and gravitational redshifting on profile of the
Fe K$\alpha$ line emitted by an accretion disk around a black hole.
Image Credit: Fabian et al., Copyright: PASP, 112, 1145 (2000).}
\label{fig07}
\end{figure}

\citet{chen89b} and \citet{chen89a} developed an approach for
calculating the intensities and profiles of the lines emitted from a
Keplerian accretion disk in which the relativistic effects are
included in the weak-field limit. Rigorous relativistic calculations
of the Fe K$\alpha$ line profile emitted from a geometrically thin
accretion disk around a stationary black hole, i.e. in the case of
Schwarzschild metric have been carried out by e.g.
\citet{fab89,ste90,mat92}. The relativistic effects on the continuum
emission from an accretion disk around a rotating black hole (i.e.
for Kerr metric) were first calculated by \citet{cun75}, while the
line profiles in such a case were simulated by \citet{laor91}.
\citet{fab00} studied the influence of Doppler effect and
gravitational redshift on the shape of Fe K$\alpha$ line emitted by
an accretion disk around a black hole and found that, as a result,
the line profile appears to be broadened and skewed (see Fig.
\ref{fig07}).

\begin{figure*}[ht!]
\centering
\includegraphics[width=0.75\textwidth]{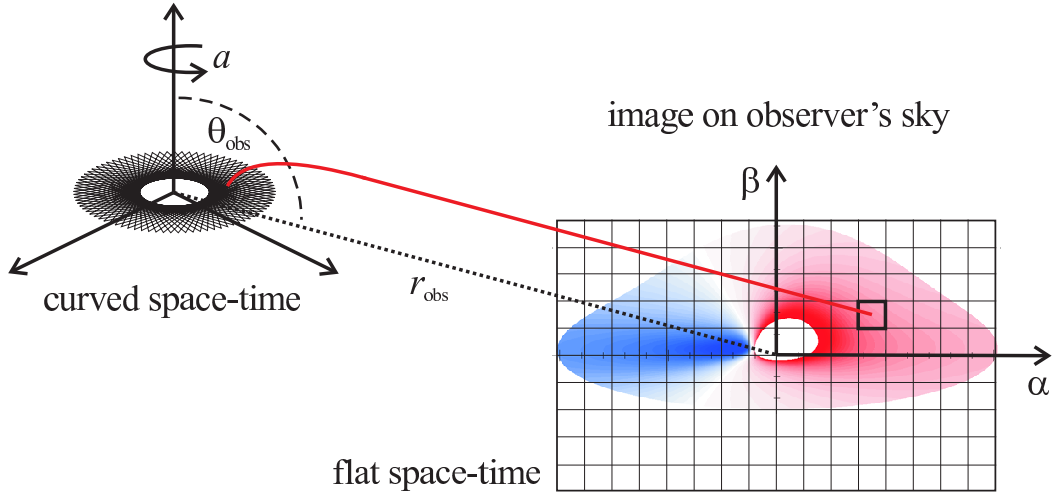}
\caption{Schematic illustration of the ray-tracing method in the
Kerr metric, showing a light ray emitted from some radius of
accretion disk in coordinate system defined by a rotating black hole
with angular momentum $a$, and observed at a pixel with coordinates
(impact parameters) $\alpha,\beta$ on the disk image in the
observer's reference frame. Image Credit: Jovanovi\'{c} \&
Popovi\'{c}, Copyright: Black Holes and Galaxy Formation, Nova
Science Publishers Inc, Hauppauge NY, USA, 249-294 (2009).}
\label{fig08}
\end{figure*}

The emission from accretion disk can be analyzed also by numerical
simulations based on ray-tracing method in Kerr metric
\citep{Bao94,Bromley97,Fanton97,Cadez98}, taking into account only
photon trajectories reaching the observer's sky plane. In this
method one divides the image of the disk on the observer's sky into
a number of small elements (pixels). For each pixel, the photon
trajectory is traced backward from the observer by following the
geodesics in a Kerr space-time, until it crosses the plane of the
disk (see Fig. \ref{fig08}). Then, the flux density of the radiation
emitted by the disk at that point, as well as the redshift factor of
the photon are calculated. In that way, one can obtain the color
images of the accretion disk which a distant observer would see by a
high resolution telescope. The simulated line profiles can be
calculated taking into account the intensities and received photon
energies of all pixels of the corresponding disk image.
\citet{Cadez98} developed a variant of ray-tracing method based on
the pseudo-analytical integration of the geodesic equations which
describe the photon trajectories in the Kerr metric.

\begin{figure*}[ht!]
\centering
\includegraphics[height=0.75\textwidth,angle=270]{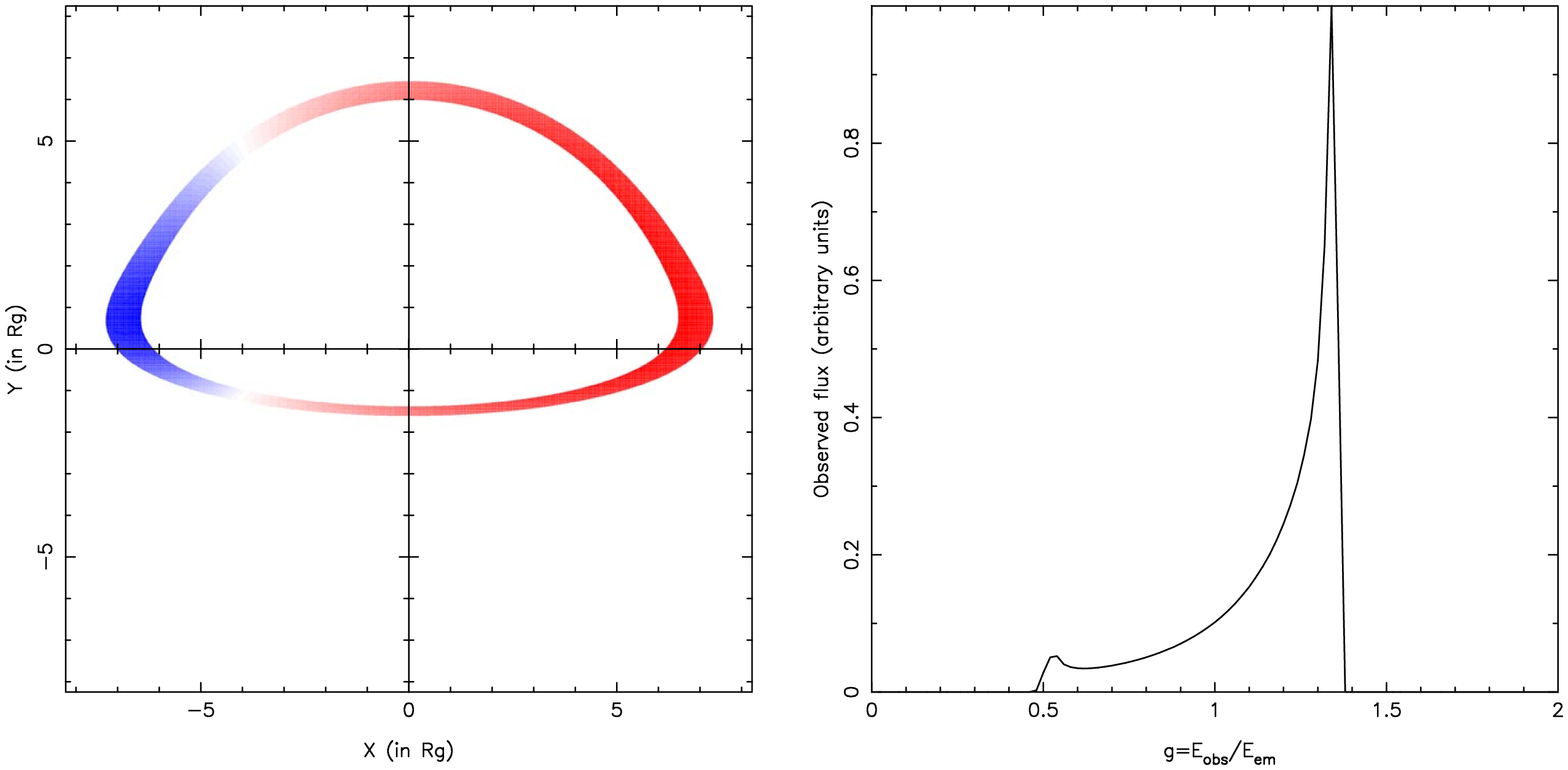} \\
\includegraphics[height=0.75\textwidth,angle=270]{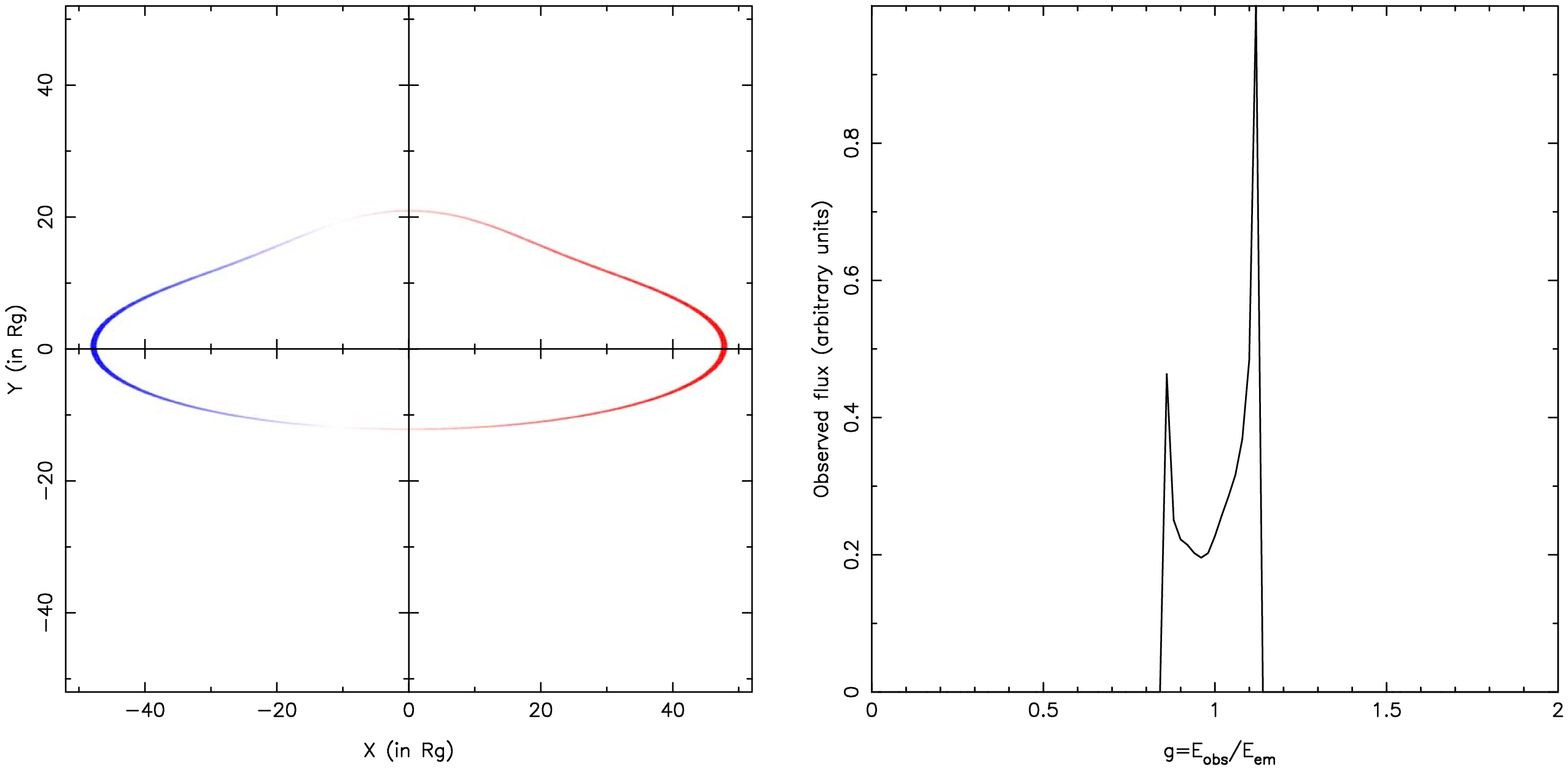}
\caption{\textit{Left:} illustrations of the Fe K$\alpha$ line
emitting region in form of narrow ring with width $=1 R_{g}$,
extending from: $R_{in}=6$ R$_{g}$ to $R_{out}=7$ R$_{g}$ (top) and
$R_{in}=50$ R$_{g}$ to $R_{out}=51$ R$_{g}$ (bottom).
\textit{Right:} the corresponding Fe K$\alpha$ line profiles. Image
Credit: Jovanovi\'{c} \& Popovi\'{c}, Copyright: Fortschr. Phys.,
56, 456 (2008).} \label{fig09}
\end{figure*}

The complex profile of the Fe K$\alpha$ line depends on different
parameters of the accretion disk and central black hole \citep[see
e.g.][]{fab00}. In order to study the size of the Fe
K$\alpha$ line emitting region, as well as its location in the disk,
one can assume that the line is emitted from a region in form of a
narrow ring. For example, \citet{jov08} assumed a line emitting
region with width equals to $1\ R_{g}$, located between: a)
$R_{in}=6$ R$_{g}$ and $R_{out}=7$ R$_{g}$ and b) $R_{in}=50$ R$_{g}$
and $R_{out}=51$
R$_{g}$. These two cases are presented in Fig. \ref{fig09}. From
Fig. \ref{fig09} one can see how the Fe K$\alpha$ line profile is
changing as the function of distance from central black hole. When
the line emitters are located at the lower radii of the disk, i.e.
closer to the central black hole, they rotate faster and the line is
broader and more asymmetric (see Fig. \ref{fig09} top-right). If the
line emission is originating at larger distances from the black
hole, its emitting material is rotating slower and therefore the
line becomes narrower and more symmetric (see Fig. \ref{fig09}
bottom-right). In majority of AGN, where the broad Fe K$\alpha$ line
is observed, its profile is more similar to the modeled profile as
obtained under assumption that the line emitters are located close
to the central black hole \citep{Tanaka95,Nandra07,jov08}.

\begin{figure*}[ht!]
\centering
\includegraphics[width=0.75\textwidth]{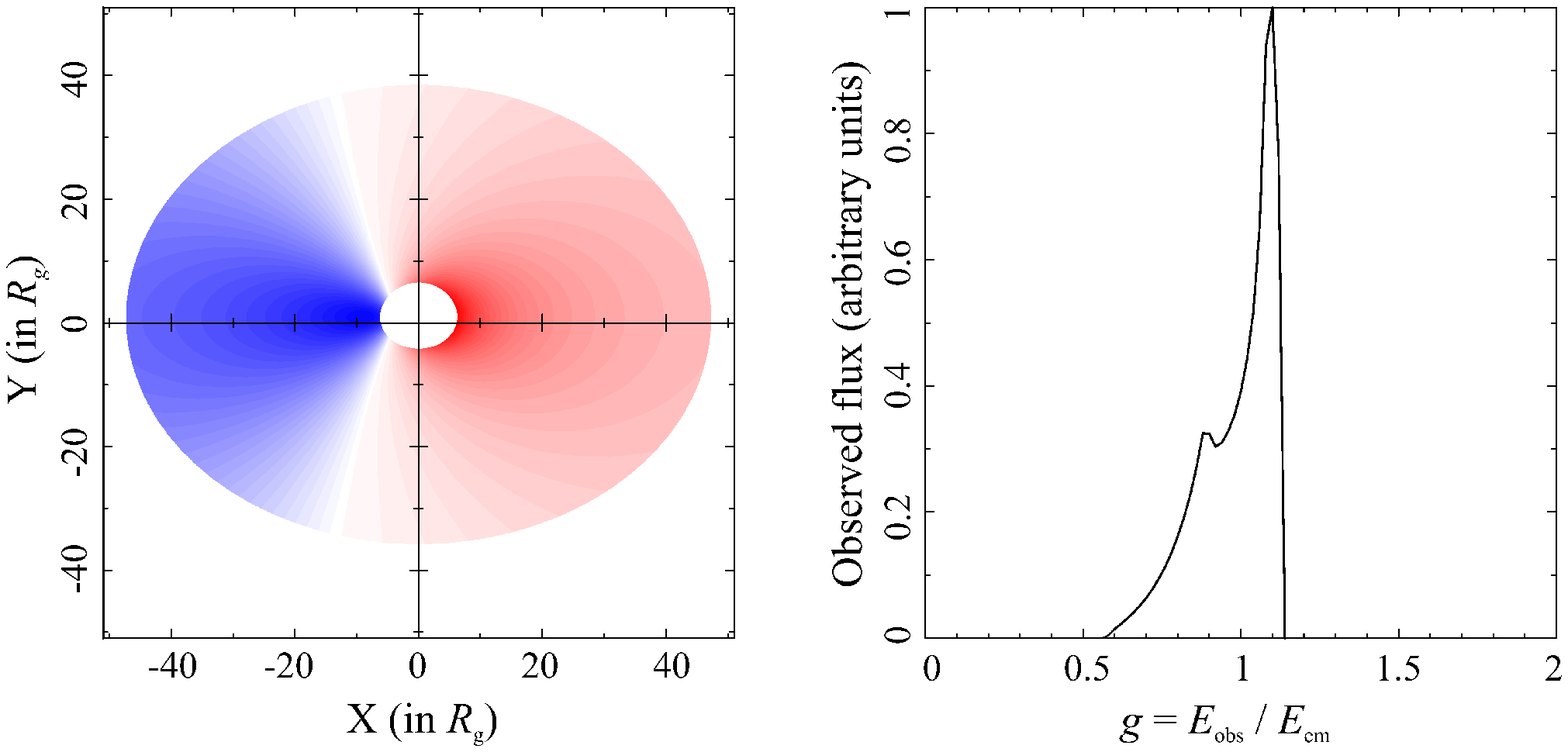} \\
\includegraphics[width=0.75\textwidth]{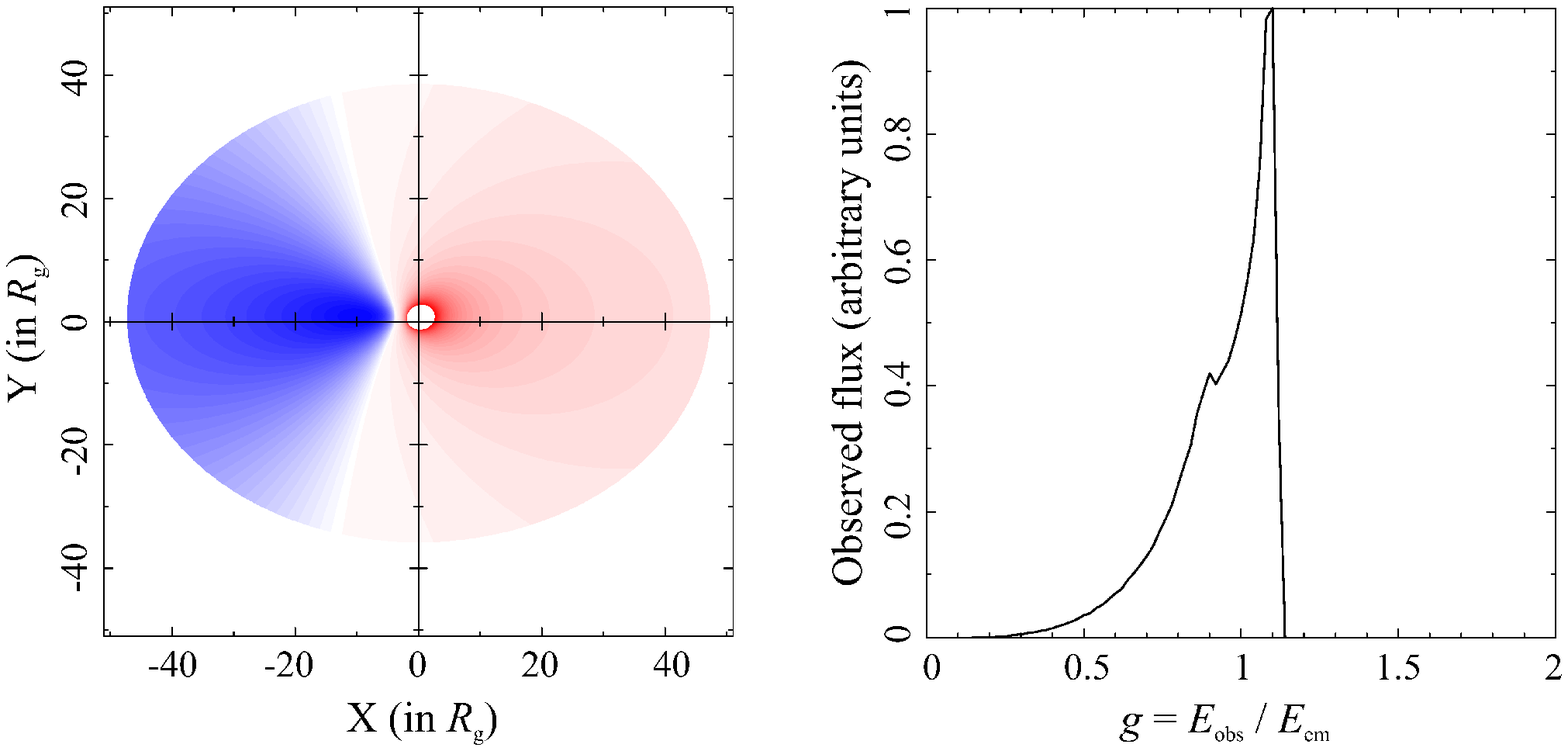}
\caption{Illustrations of an accretion disk (left panels) and the
corresponding Fe K$\alpha$ line profiles (right panels) for disk
inclination $i$ = 40$^\circ$ in the Kerr metric for angular momentum
$a$ = 0.1 (\emph{top}) and $a$ = 0.998 (\emph{bottom}). Image
Credit: Jovanovi\'{c} et al., Copyright: Baltic Astronomy (2011).}
\label{fig10}
\end{figure*}

\begin{figure}[hb!]
\centering
\includegraphics[width=0.9\columnwidth]{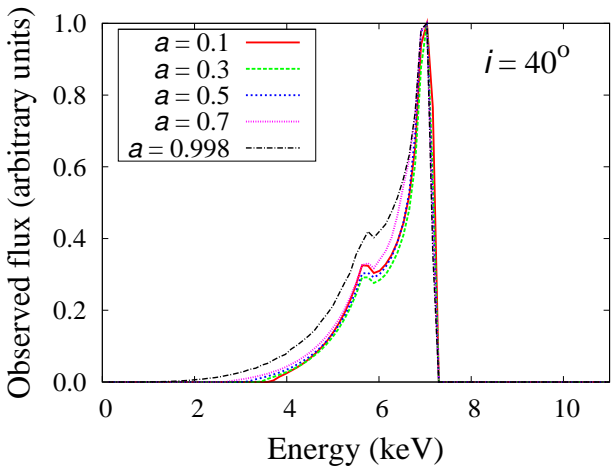}
\caption{Modeled Fe K$\alpha$ spectral line profiles for several
values of angular momentum parameters $a$, and for inclination angle
$i$ = 40$^\circ$. Image Credit: Jovanovi\'{c} et al., Copyright: Baltic Astronomy (2011).}
\label{fig11}
\end{figure}

Angular momentum or spin of the central supermassive black hole of
AGN is a property of the space-time metric and different
techniques for its estimation are developed
\citep[see e.g.][]{cz11,daly11}. To study the effects of
black hole spin on the shape of the Fe K$\alpha$ line we performed
simulations of disk emission based on ray-tracing method in Kerr
metric. Simulated images of an accretion disk around a supermassive
black hole and the corresponding Fe K$\alpha$ line profiles for two
different values of its spin are presented in Fig. \ref{fig10}
\citep[see][for more details]{jov11}. In these simulations, the disk
extends from the radius of marginally stable orbit ($R_{ms}$) to 40\
$R_g$, where $R_g$ is the gravitational radius. As one can see from
Fig. \ref{fig10}, $R_{ms}$ strongly depends on black hole spin,
which consequently also affects the profile of the line originating
in the innermost parts of the disk \citep[see also][]{bren06}.
Simulated Fe K$\alpha$ line profiles for several values of black hole
spin are
presented in Fig. \ref{fig11} \citep{jov11}. It can be seen from
Fig. \ref{fig11} that the black hole spin especially affects the red
wing of the Fe K$\alpha$ line, and that this wing is more extended
towards lower energies for higher values of the spin
\citep[see][]{rey03,jov08,jov11}. At the same time, the line becomes
wider and its red peak brighter.

\begin{figure*}[ht!]
\centering
\includegraphics[height=0.75\textwidth,angle=270]{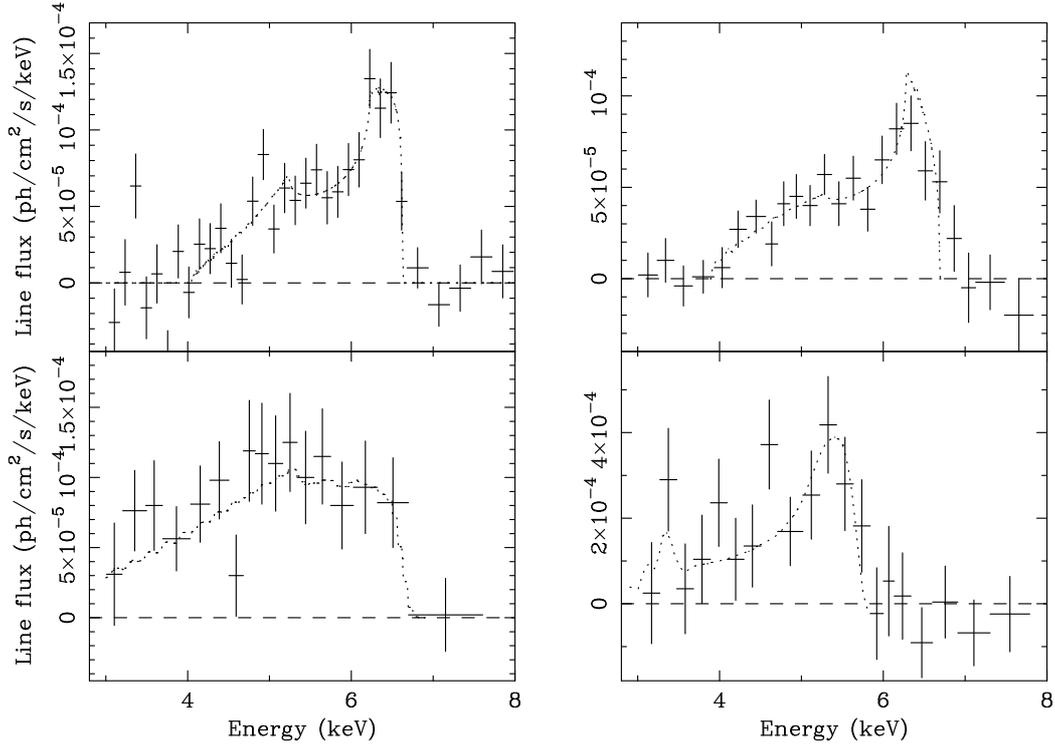}
\caption{Time-averaged (upper panels) and peculiar line profiles
(lower panels) of the Fe K$\alpha$ line in the case of MCG --6-30-15
observed by \emph{ASCA} in 1994 (left panels) and 1997 (right
panels). Image Credit: Fabian et al., Copyright: PASP, 112, 1145
(2000).} \label{fig12}
\end{figure*}

\subsection{Variability of the Fe K$\alpha$ line}

One of the important features of the Fe K$\alpha$ line is
variability of both, its shape and intensity \citep[see e.g.][and
references therein]{bhan10}. Fig. \ref{fig12} shows
one of the most illustrative examples for such variations which was
found in \emph{ASCA} observations of MCG--6-30-15 during 1994 and
1997 \citep{fab00}. As it can be seen from Fig. \ref{fig12}, in the
1994 observation, a very broad profile with a pronounced red wing is
seen during a period of deep minimum of the light curve (lower left
panel), compared to the time-averaged line profile shown in the
upper panel. In contrast, during a sharp flare in the 1997
observation, the whole line emission is shifted to energies below 6
keV, and there is no significant emission at the rest line energy of
6.4 keV (lower right panel). Both peculiar line shapes can be
explained by large gravitational redshift in small radii on the
accretion disk \cite{fab00}.

Variations of double-peaked emission lines are mainly attributed to
physical changes in the accretion disk, but in case of some AGN they
cannot be explained by a simple, circular accretion disk model, and
thus the need for asymmetries in the accretion disk arose. Such
asymmetries are usually introduced in form of a precessing
elliptical disk \citep{er95}, a circular disk with a spiral arms
\citep{lew10} or with rotating bright spots \citep{fl08}.

\begin{figure}[hb!]
\centering
\includegraphics[height=\columnwidth,angle=270]{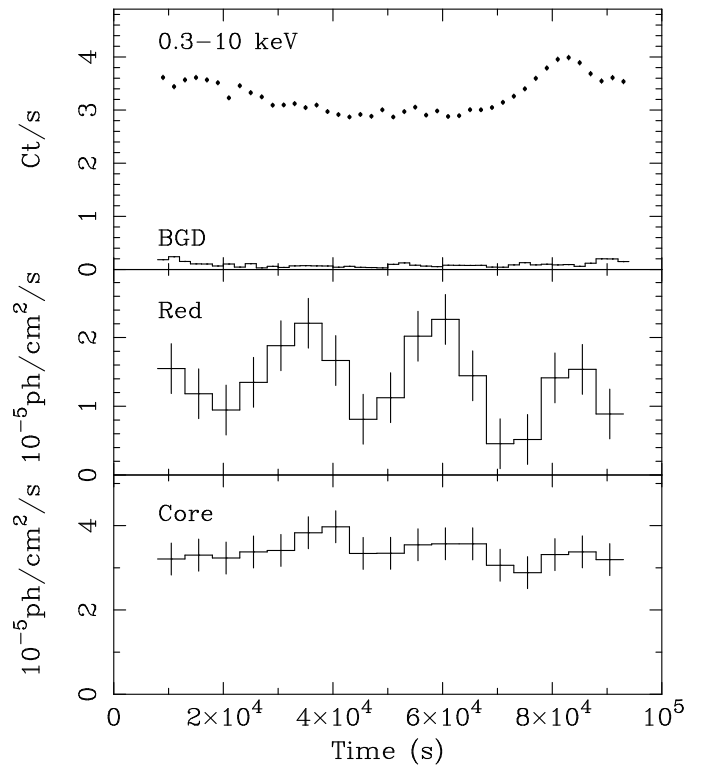}
\caption{Light curves of Seyfert galaxy NGC 3516 \citep{Iwasawa04}
for: the 0.3 -- 10 keV band (top), the Fe K$\alpha$ line red feature
(middle) and the 6.4 keV line core (bottom).
Image Credit: Iwasawa et al., Copyright: MNRAS, 355, 1073 (2004).}
\label{fig13}
\end{figure}

\begin{figure*}[ht!]
\centering
\includegraphics[width=0.74\textwidth]{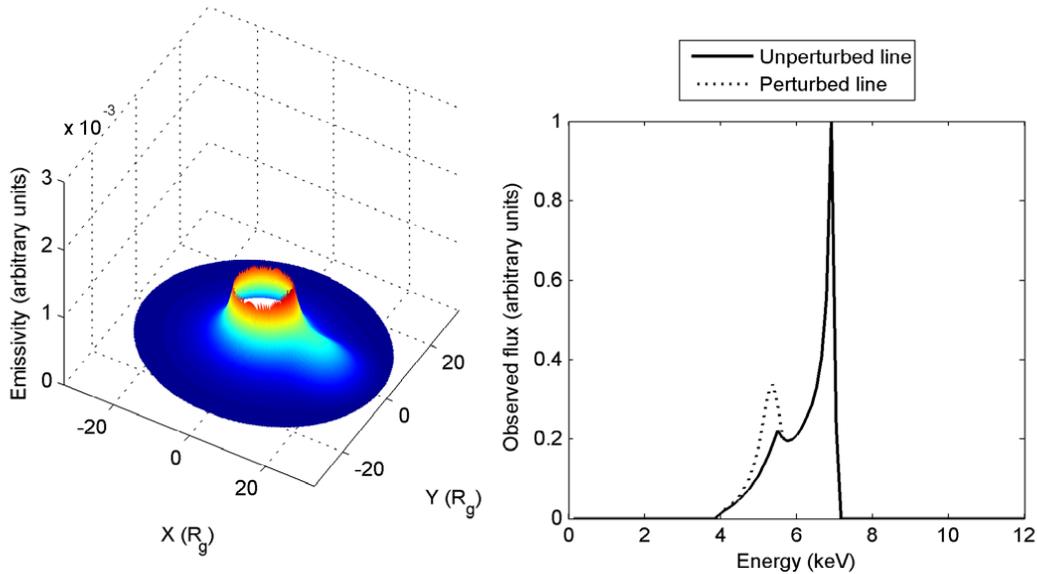}
\caption{\textit{Left:} perturbed emissivity of an accretion disk in
Schwarzschild metric \citep{jov09}. \textit{Right:} the
corresponding perturbed (dashed line) and unperturbed (solid line)
Fe K$\alpha$ line profiles. Image Credit: Jovanovi\'{c} \&
Popovi\'{c}, Copyright: Black Holes and Galaxy Formation, Nova
Science Publishers Inc, Hauppauge NY, USA, 249-294 (2009).}
\label{fig14}
\end{figure*}

\citet{Iwasawa04} detected a variable ''red'' feature of the Fe
K$\alpha$ line at 6.1 keV (in addition to the stable 6.4 keV line
core) in X-ray spectrum of Seyfert galaxy NGC 3516, observed by
\emph{XMM-Newton} satellite (see Fig. \ref{fig13}). This feature
varies systematically in the flux at intervals of 25 ks and in
energy between 5.7 and 6.5 keV. \citet{Iwasawa04} found that the
spectral evolution of the ''red'' feature agrees well with
hypothesis of an orbiting spot in the accretion disk.

Appearances of such bright spots could be described by perturbations
in accretion disk emissivity \citep{jov10} and could be caused by
several physical mechanisms, such as: disk self-gravity, baroclinic
vorticity, disk-star collisions, tidal disruptions of stars by
central black hole and fragmented spiral arms of the disk \citep[see
e.g.][and references therein]{jov10}. All these phenomena have
different occurrence frequencies, durations, timescales, strengths,
proportions and other characteristics. Especially, perturbations of
accretion disk emissivity in form of flares with high amplitudes are
of great significance because they could provide information on
accretion physics under extreme conditions. Such flares with the
highest amplitudes are usually interpreted in terms of tidal
disruptions of stars by supermassive black holes \citep[see
e.g.][and references therein]{kom08}. Stars approaching a SMBH will
be tidally disrupted once the tidal forces of the SMBH exceed the
star's self-gravity, and part of the stellar debris will be
accreted, producing a luminous flare of radiation which lasts on the
timescale of months to years. Although, frequency of such events in
a typical elliptical galaxy is very low, between $10^{-5}$ and
$10^{-4}$ per year \citep[see e.g.][and references therein]{jov10},
recently \citet{kom08} reported the discovery of an X-ray outburst
of large amplitude in the galaxy SDSS J095209.56+214313.3 which was
probably caused by the tidal disruption of a star by a supermassive
black hole. A simulation of such perturbed emissivity of an accretion
disk in Schwarzschild metric, as well as the corresponding perturbed
and unperturbed Fe K$\alpha$ line profiles, are presented in the left
and right panels of Fig. \ref{fig14}, respectively \citep{jov09}.

Most AGN contain energetic outﬂows of ionized gas, emerging from
their accretion disks at relativistic speeds, which imprint multiple
broad absorption features blueward of the emission lines \citep[see
e.g.][]{chart02}. These additional absorption components, arising
from outflowing winds, may distort the energy region near the Fe
K$\alpha$ line. Recently, \citet{chart09} confirmed the presence of
X-ray broad absorption lines from the quasar APM 08279+5255, varying
on short timescales of several days which implies a source
size-scale of $\sim 10\ R_g$ \citep[see also][]{bran09}.
\citet{Done07} found an evidence for a P Cygni profile of the Fe
K$\alpha$ line in narrow line Seyfert 1 galaxies and showed that a
sharp drop in a such profile at $\sim 7$ keV results from
absorption/scattering/emission of the iron K$\alpha$ line in the
outflowing wind.

In addition to intrinsic causes such as a disk instability, the Fe
K$\alpha$ line variability could also be induced by some external
effects, such as absorption by X-ray absorbers or gravitational
microlensing \citep{jov09}. Observations of so-called Low Ionization
Broad Absorption Line quasars \citep[e.g. Mrk 231][]{Braito04} and
\citep[H 1413+117][]{Chartas07} confirmed the presence of X-ray
absorbers in these objects. \citet{Wang01} detected an absorption
line at 5.8 keV in nearby Seyfert 1.5 galaxy NGC 4151 and a variable
absorption line at the same energy has been discovered by
\citet{Nandra99} in NGC 3516. It was interpreted as a Fe K resonant
absorption line, redshifted either by infalling absorbing material
or by strong gravity in the vicinity of the black hole. A model of
the Fe K$\alpha$ line absorbing/obscuring regions in form of
absorbing medium comprised of cold absorbing cloudlets was developed
by \citet{Fuerst04}. Another model, which assumes that absorbing
region is composed of a number of individual spherical absorbing
clouds, was developed by \citet{jov07}.

\cleardoublepage

\begin{figure*}[ht!]
\centering
\includegraphics[width=\textwidth]{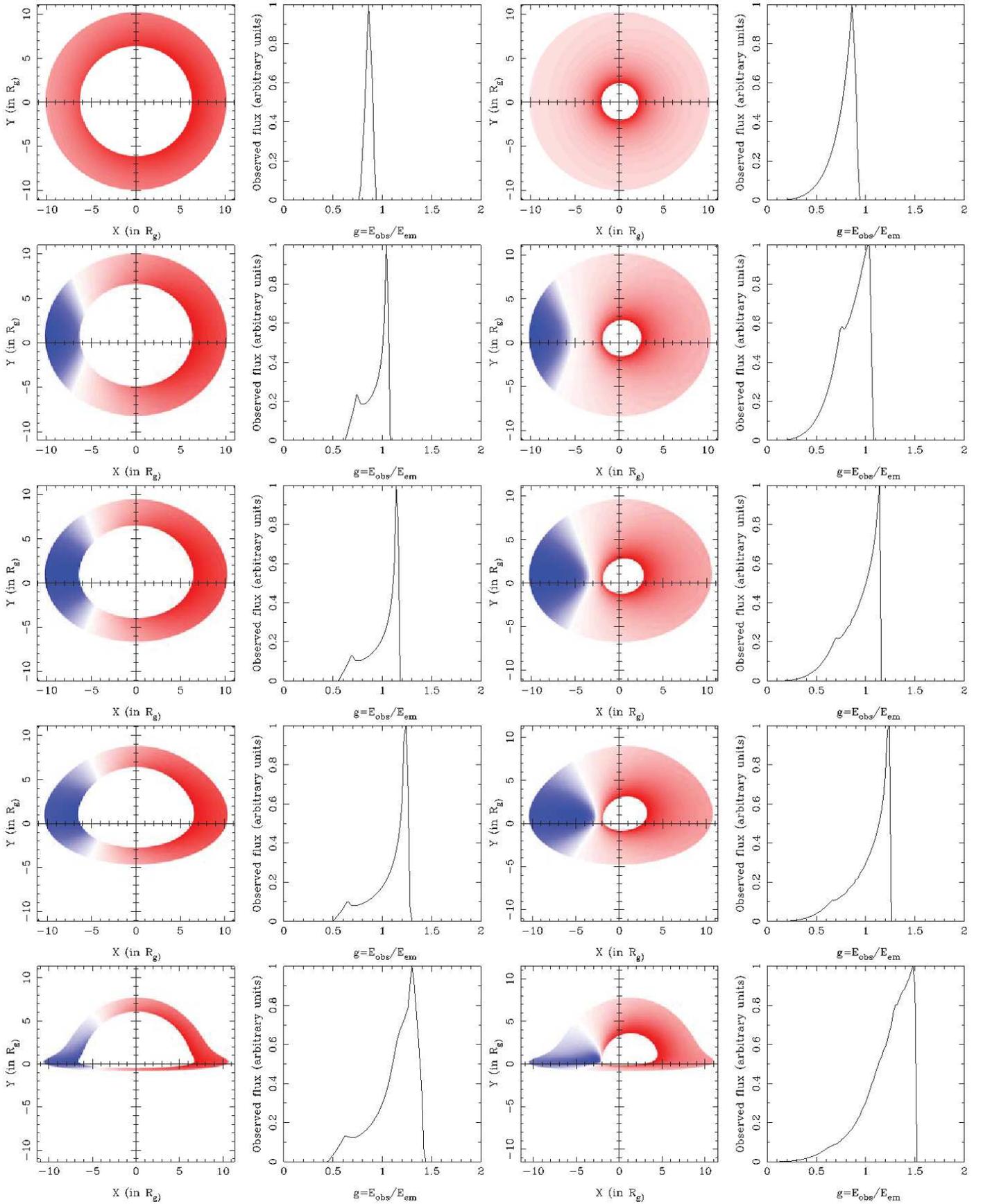}
\caption{\textit{Left:} Simulated images of accretion disk in
Schwarzschild metric and the corresponding Fe K$\alpha$ line
profiles. \textit{Right:} The same, but in Kerr metric for maximally
rotating black hole, i.e. for $a=0.998$. Disk inclination takes the
following values from top to bottom panels, respectively:
$i=5^\circ,\,30^\circ,\,45^\circ,\,60^\circ,\,85^\circ$.}
\label{fig15}
\end{figure*}

\cleardoublepage

\citet{pop01} found that a compact gravitational microlens could
induce noticeable changes in the Fe K$\alpha$ line profile when the
Einstein ring radius associated with the microlens is comparable to
the size of the accretion disk. Also, the X-ray continuum could
experience significant amplification during such microlensing events
\citep{pop06b}. \citet{Zakharov04} found that cosmologically
distributed gravitational microlenses could significantly contribute
to the X-ray variability of high-redshifted ($z>2$) quasars. Indeed,
microlensing of the Fe K$\alpha$ line has been reported at least in
three lensed quasars with multiple images: MG J0414+0534
\citep{Chartas02}, QSO 2237+0305 \citep{Dai03}, and H 1413+117
\citep{Oshima01,Chartas04}. \citet{Chartas02} detected an increase
of the Fe K$\alpha$ equivalent width in the image B of MG J0414+0534
which was not followed by the continuum and explained this behavior
by assumption that the thermal emission region of the disk and the
Compton up-scattered emission region of the hard X-ray source lie
within smaller radii than the iron-line reprocessing region.
\citet{Dai03} also measured amplification of the Fe K$\alpha$ line
in component A of QSO 2237+0305. For more detailed discussion about
observational and theoretical investigations of the Fe K$\alpha$
line variability due to gravitational microlensing see e.g.
\citet{jov05,jov06,jov08b} and \citet{pop03a,pop03b,pop06}, as well
as references therein.

\section{Discussion}

In order to demonstrate the consequences on the Fe K$\alpha$ line of
some general relativistic and strong gravitational effects, which
are significant only near the marginally stable orbit, we performed
ray-tracing simulations \citep[using approach proposed
by][]{Cadez98} of a disk emission from its innermost regions. For
that purpose we analyzed several cases for disk inclination and
spin, assuming that the disk extends from the $R_{ms}$ to 10 $R_g$,
which is the region where the line is most likely produced in most
AGN \citep[see e.g][]{ball05}.

The obtained results are presented in Fig. \ref{fig15}. As one can
see from the Fig. \ref{fig15}, for all studied inclinations there
are significant differences between the simulated line profiles in
the case of non-rotating and maximally rotating black hole. In the
latter case the line profiles are generally wider and more extended
towards lower observed energies (see also Fig. \ref{fig11}), due to
fact that the disk extends from the radius of marginally stable
orbit, which is much smaller in this case \citep{jov11}. This
effect, when observed, gives opportunity to measure black hole spin,
and thus to have insight into space-time geometry in its vicinity.
At the same time, as a consequence of gravitational redshift, the
''blue'' peak is more smeared than the ''red'' one in the case of
maximally rotating black hole. Therefore, comparisons between the
observed and simulated Fe K$\alpha$ line profiles enables us to
investigate the effects of strong gravitational field predicted by
General Relativity. However, since the spectrum ''reflected'' from
the disk may contain not only a single iron line, but also many
fluorescent lines and a continuum reflection component \citep[see
e.g.][]{fab06}, accurate modeling of the underlying reflected and
direct components is crucial in using the Fe K$\alpha$ line as a
diagnostic of General Relativity in the strong field regime.

Except the black hole spin, the disk inclination also have
significant influence on the line profile, as can be seen from Fig.
\ref{fig15}. For small inclinations, i.e. in the case of almost
face-on disk, the single peak profiles of the Fe K$\alpha$ line are
obtained. For inclinations about $i\approx 30^\circ$, the faint
''red'' peak is the most emphasized. It is interesting that this
value is close to the averaged value of $i=35^\circ$, estimated by
\cite{Nandra97} from the study of the Fe K$\alpha$ line profiles of
18 Seyfert 1 galaxies.

\section{Conclusions}

The results of previously mentioned observational and theoretical
studies indicate that the broad Fe K$\alpha$ line which originates
in vicinity of the supermassive black holes is a powerful tool for
studying their properties, space-time geometry (metric) in their
vicinity, their accretion physics, probing the effects of their
strong gravitational fields, and for testing the certain predictions
of General Relativity. However, some important scientific issues
still need to be addressed, such as decoupling the narrow and broad
Fe K$\alpha$ line components and explaining their different
variabilities \citep[see e.g][]{sul98}. This will be one of the
crucial steps in determining the real importance of relativistic
effects on the observed Fe K$\alpha$ line profiles.

\bigskip\noindent\textit{Acknowledgements.} This work is part
of the project 176003 ''Gravitation and the Large Scale Structure of
the Universe`` supported by Ministry of Education and Science of the
Republic of Serbia. It was presented as an invited talk at special
workshop ''Spectral lines  and super-massive black holes'' held on
June 10, 2011 as a part of activity in the frame of COST action
0905 ''Black holes in a violent universe'', during the ''8th
Serbian Conference on Spectral Line Shapes in
Astrophysics''.

\end{document}